\documentclass[prap,twocolumn,showpacs,floatfix]{revtex4-1}
\usepackage{times}
\usepackage{bm}
\usepackage[dvips]{graphicx}
\usepackage{amsmath}
\usepackage{natbib}
\usepackage{color}
\usepackage{epstopdf}

\begin{document}

\title{Effects of hole self-trapping by polarons on transport and negative bias illumination stress in amorphous-IGZO}

\author{A.~de~Jamblinne~de~Meux$^{1,2}$, G. Pourtois$^{2,3}$,\\ J. Genoe$^{1,2}$ and P. Heremans$^{1,2}$}

\affiliation{
$^1$ KU Leuven,  ESAT, B-3001 Leuven, Belgium
$^2$ IMEC, Kapeldreef 75, B-3001 Leuven, Belgium
$^3$ Department of Chemistry, Plasmant Research Group, University of Antwerp, B-2610 Wilrijk-Antwerp, Belgium.
}

\date{\today}

\begin{abstract}

The effects of hole injection in amorphous-IGZO is analyzed by means of first-principles calculations. The injection of holes in the valence band tail states leads to their capture as a polaron, with high self-trapping energies (from 0.44 to 1.15 eV). Once formed, they mediate the formation of peroxides and remain localized close to the hole injection source due to the presence of a large diffusion energy barrier (of at least 0.6eV). Their diffusion mechanism can be mediated by the presence of hydrogen. The capture of these holes is correlated with the low off-current observed for a-IGZO transistors, as well as, with the difficulty to obtain a p-type conductivity. The results further support the formation of peroxides as being the root cause of Negative bias illumination stress (NBIS). The strong self-trapping substantially reduces the injection of holes from the contact and limits the creation of peroxides from a direct hole injection. In presence of light, the concentration of holes substantially rises and mediates the creation of peroxides, responsible for NBIS.

\end{abstract} 

\maketitle

\section{Introduction} 

Amorphous Indium-Gallium-Zinc-oxide (a-IGZO) has become the reference high mobility amorphous oxide semiconductor. Compared to a-Si, it combines an improved mobility, a lower off-current and a better stability \cite{Kamiya2010,YeonKwon2015}. Unfortunately, its stability can be compromised by a combination of bias stress and light\cite{Jeong2013b}. This phenomenon is called Negative Bias Illumination Stress (NBIS): large shifts of the transistor transfer curve are observed whenever a negative bias is applied together with blue or near UV illumination. Several sources of NBIS have been proposed. For instance, Jeong et al.\cite{Jeong2013b} review three different mechanisms contributing to NBIS: Interaction with the ambient, the charge trapping at the interfaces, and in the bulk semi-conductor. 

The influence of the ambient is typically solved by the application of a proper passivation of the channel \cite{Jeong2013b,Nomura2011b}. Similarly, the charge trapping at interfaces is significantly reduced with the development of high quality interfaces \cite{Nomura2011b}. However, even in these devices, NBIS is found to induce a considerable (several volt) and quasi-permanent (beyond a few days) shift of the transistor characteristics when the transistor is illuminated with photon energy larger than $>$ 2.9eV \cite{Nomura2011b}. The mechanism which is usually invoked to explain NBIS is the ionization of neutral oxygen vacancies (Vo) into doubly ionized ones (Vo2+) \cite{Ryu2010, Noh2011d, Park2012,Migliorato2012, Lee2015}. In this model, Vo are deep donors, while Vo2+ are shallow ones. Under illumination, the Vo sites become ionized and positively charged (Vo2+). This transition from a deep to a shallow donor dopes the semiconductor and could explain the shift of the transistor transfer curve. It is then assumed that a high ionization barrier exists between the two states (Vo and Vo2+), forbidding any recovery process to occur.

The oxygen vacancy theory originates from early DFT calculations\cite{Kamiya2009e}, which associated the presence of deep states close to the top of the valence band to oxygen vacancies, and showed that electrons could be trapped by neutral Vo upon the formation of a metal-metal bond. Further studies have however challenged this mechanism and the association of Vo to the signature of deep states located close to the tops of the valence band \cite{me2, Nomura2011c,Korner2013d, Sallis2014, Robertson2014}.

The strong negative impact of hydrogen on NBIS \cite{Jeong2013a, Kang2015} also contradicts the model based on oxygen vacancies. Indeed Noh et al. \cite{Noh2013} predicted through DFT simulations that hydrogen should passivate the vacancies and lessen the NBIS. However, the opposite effect is observed: the NBIS stability deteriorates with the increase of hydrogen content\cite{Jeong2013a, Kang2015}. Further a-IGZO films with a low electron concentration (~10$^{15}$cm$^{-3}$) can still contain an impressive amount of hydrogen (10$^{20}$cm$^{-3}$)\cite{Nomura2013}. This low electron concentration was explained by the incorporation of an excess of oxygen which acts as electron acceptors and hence compensates the impact of shallow hydrogen donors \cite{Nomura2013,Robertson2014,Han2015}. 

An alternative theory for NBIS is based on the creation of peroxides \cite{Nahm2012,Jeong2013b,Robertson2014,Han2015}. These can form spontaneously in presence of a fermi-level deep in the gap (i.e. upon the application of a strong bias stress or light illumination), and are meta-stable whenever the Fermi-level is moved in the conduction band. In amorphous IGZO, the oxygen atoms are bonded to metals and acquire a -2 oxidation state. If two oxygens are bonded together, they form a peroxide. The electrons involved in the newly formed covalent bond are shared between the oxygen sites instead of contributing to the formation of bonds with the surrounding cations. This effectively "frees" two electrons from the cation and acts as a donor source, doping the material and shifting the transistor curve.

While the formation of peroxides may explain NBIS, the exact mechanism remains unclear. Indeed, both light and negative bias stress alone should be able to generate peroxides but only the combination of the two generates NBIS. Further, hydrogen has a strong effect on NBIS\cite{Jeong2013a, Kang2015} and the reasons why the formation of a peroxide would be impacted so strongly by the presence hydrogen are unclear. Through a systematic study of the impact of holes in a-IGZO, we show that polarons form and trap holes in a-IGZO. These mediate the formation of peroxides and clarify the role of the combination of light excitation and negative bias in NBIS. We further provide an explanation of the role of hydrogen in this mechanism.

The self-trapping of these holes also provides an explanation of the origin of the low off-current observed in a-IGZO transistors \cite{Kamiya2010, Nomura2011a}.

\section{Methodology}

In contrast to crystalline materials, models of amorphous structures are not unique and small variations in their atomic topology can lead to a modulation of the response to hole(s) injection. The modeling of these systems is hence very challenging due to the statistical nature of the morphology and requires the use of atomistic models with sufficiently large dimensions \cite{me2}. In order to capture this effect, we generated eleven different atomistic models with a fixed $\rm InGaZnO_4$ stoichiometry. The first ten models own 105 atoms in their unit cell (which we numbered for the sake of the discussion from 1 to 10 in the following), while the last one has been extended to 315 atoms (labeled as being structure 11).  

The simulations were performed within the framework of the density functional theory (DFT) with a mix of Gaussian and plane wave basis, as implemented in the CP2K package \cite{cp2k}. Core electrons were modeled by GTH pseudo-potentials \cite{PhysRevB.54.1703,PhysRevB.58.3641,krack2005pseudopotentials} with a planewave cutoff of 600Ry and a relative cutoff of 50Ry  integrated with 5 sub-grids.  Molecular dynamics were performed with a single $\rm \zeta$ quality Gaussian basis set \cite{hartwigsen1998relativistic} for the metals and double $\rm \zeta$ valence polarized basis\cite{hartwigsen1998relativistic} for the oxygen together with the exchange-functional of Perdew, Burke and Ernzerhof (PBE)\cite{PhysRevLett.77.3865} using the procedure described in reference \cite{Meux2015}. The use of single $\rm \zeta$ quality basis for the metal allows for a significant reduction of the computational cost, while maintaining the structural and electronic properties of the material unaffected\cite{me1}.

To account for the impact of the electronic correlation effects on the structural and electronic properties of a-IGZO and to alleviate the notably strongly underestimated band gap of a-IGZO within PBE, all structural relaxations and electronic properties were obtained using a hybrid PBE0 exchange correlation functional \cite{VandeVondele2005103,Guidon2009}. The evaluation of the Hartree-Fock exchange was realized with a truncated Coulomb operator with a cutoff radius of 3$\rm \AA$ \cite{VandeVondele2005103,Guidon2009}. We observed that for the 315 atoms structure, this cut-off leads to similar results (with a variation of 0.04eV/atom on the total energy) than with the maximum cut-off of 7.8$\rm \AA$ but for a more affordable computational cost. A further reduction of the computational cost was achieved through the use of the auxiliary density matrix method \cite{doi:10.1021/ct1002225}. The auxiliary basis set used consists of an uncontracted basis with eleven Gaussian exponents for the metal and nine ones for the oxygen. All structures were fully relaxed with the hybrid exchange-functional until the resulting atomic forces are lower than 0.001Ry/Bohr. The volume of the models was relaxed until a threshold for the internal pressure of 1KBar is reached. Systems containing one hole (i.e an odd number of electrons), were treated using a spin polarization formalism.   

All the models were generated using a melt and quenching method\cite{Meux2015}. The ten 105 atoms models originate from the work detailed in reference \cite{Meux2015} and were relaxed with the PBE0 hybrid functional. The initial starting structure for the 315 atoms modes was built by a method similar to the seed-coordinate-anneal one proposed in reference \cite{Youn2014} and subsequently annealed and relaxed. The density of states, mean coordination and bond lengths of all models are similar as illustrated in the figures 1 and 2 of the supplemental materials.

Given the high compliance of the amorphous structures, barrier heights and their associated minimal energy paths cannot be computed with a simple linear interpolation between the initial and final configurations. Therefore, the barrier height for the creation and dissociation of the defective centers induced upon hole injection (i.e. peroxide centers, see below) were evaluated by the climbing image nudged elastic band algorithm using a spring constant of 0.01 atomic units\cite{Jonsson1997, Henkelman2000a, Henkelman2000}. The initial paths were created with an image dependent pair potential algorithm\cite{Smidstrup2014}. Due to heavy computational burden of such simulations, the auxiliary basis set was reduced to a contracted one with nine Gaussian exponents instead of eleven, which lessens the computational cost at the expense of an acceptable error on the total energy of $\rm \sim $0.05eV. Finally, the inversion participation ratio (IPR)\cite{Murphy2011} is used to quantify the degree of localization of the tail states due to the injection of holes. The IPR is defined as:

\begin{equation}
IPR_\beta = \frac{\sum_i C_{\beta i}^4}{\left[\sum_i C_{\beta i}^2 \right]^2},
\end{equation}

where $C_{\beta i}$ is the contribution of atom $i$ to the molecular orbital $\beta$. By construction, the IPR values vary between zero and one. A null value indicates that the wave-function is localized uniformly on all the atoms of the model, while a unitary value indicates that the wave-function is localized on a single atomic site. 

\section{Results} 

\begin{table}
\begin{tabular}{c|c c}
Model & 1 hole (eV) & 2 holes (eV) \\ 
\hline 
1  & 0.44  & 0.93 \\ 
2 & 0.91 & 0.99 \\ 
3 & 0.47 & 0.67 \\ 
4 & 0.53 & 2.90 \\ 
5 & 0.64 & 0.74 \\ 
6 & 0.90 & 0.26 \\ 
7 & 0.73 & 0.61 \\ 
8 & 1.15 & 3.82 \\ 
9 & 0.75 & 2.40 \\ 
10 & 0.52 & 0.52 \\ 
11  & 0.95 & 3.45 \\ 
\end{tabular} 
\caption{Relaxation energies for the injection of one or two holes in a-IGZO. The first ten models contain 105 atoms and the last one owns 315 atoms in their unit cells.}
\label{table:relaxE}
\end{table}

The injection of one or two holes into an a-IGZO model leads to a substantial atomic relaxation process. The amplitude of this relaxation is reflected by the relaxation energy (table \ref{table:relaxE}), which represents the gain in energy due to structural changes. It is by definition null if no structural changes occur. In cases where holes are injected in a defect-free structure, this quantity also corresponds to the self-trapping energy \cite{Varley2012} since the relaxation leads to the formation of a polaron that traps holes.

Despite the large spread of the relaxation energies, four out of the eleven models (labels 4, 8, 9 and 11  in table 1) stand out with particularly high relaxation energies due to the formation of a peroxide (O-O bond) whenever two holes are injected in the structure. These cases set apart, the relaxation energies in presence of one or two holes are relatively high compared to the one obtained in crystalline oxide materials, where values ranging from 0.01eV (MgO) to 0.53eV ($\rm\beta$-$\rm Ga_2O_3$) have been reported \cite{Varley2012}. Note that the differences in relaxation energy between the models arise from the structural variations intrinsic to the amorphous phase. No noticeable difference was observed between the model owning 315 atoms and the smaller ones.
 
Peroxides (O-O) have an average bond length of 1.47\AA, which clearly distinguishes them from oxygen coordinated with metal cations (O-M), for which other oxygen atoms are second neighbors at a minimal distance of about 2.45\AA from site to site. This bond length is consistent with the values obtained for peroxides in molecules such as in hydrogen peroxide (1.49\AA~\cite{abrahams1951crystal}) or dimethyl peroxide (1.457\AA~\cite{haas1984gas}). The mean coordination of the oxygen atoms in these peroxides is found to be 3.75 and is hence comparable to the one of a 'normal' oxygen site within a-IGZO (3.67 - value averaged on the 11 structures used with a coordination sphere cut-off of 2.6\AA ). This indicates that the peroxide bond replaces the oxygen-metal one within their coordination sphere. The O-O bond is on average however much smaller (1.47\AA) compared to O-M one (2.06\AA). We found that in average, the formation of a polaron has a rather weak structural impact, with negligible variations of mean bond length and coordination. However, the oxygen involved in their formation get closer to each other by an average of 0.1\AA. Note that some larger structural reorganization can also occur. For instance, in one of our model, we observed a stronger reduction of the spacing separating the two oxygen sites contributing the most to the polaron, with a reduction of their distance by 0.53\AA . Once formed, these polarons therefore mediate the formation of peroxides by bringing the oxygen atoms closer to each other.

\begin{figure*}[hbtp]
 \centering
 \includegraphics[scale=0.8]{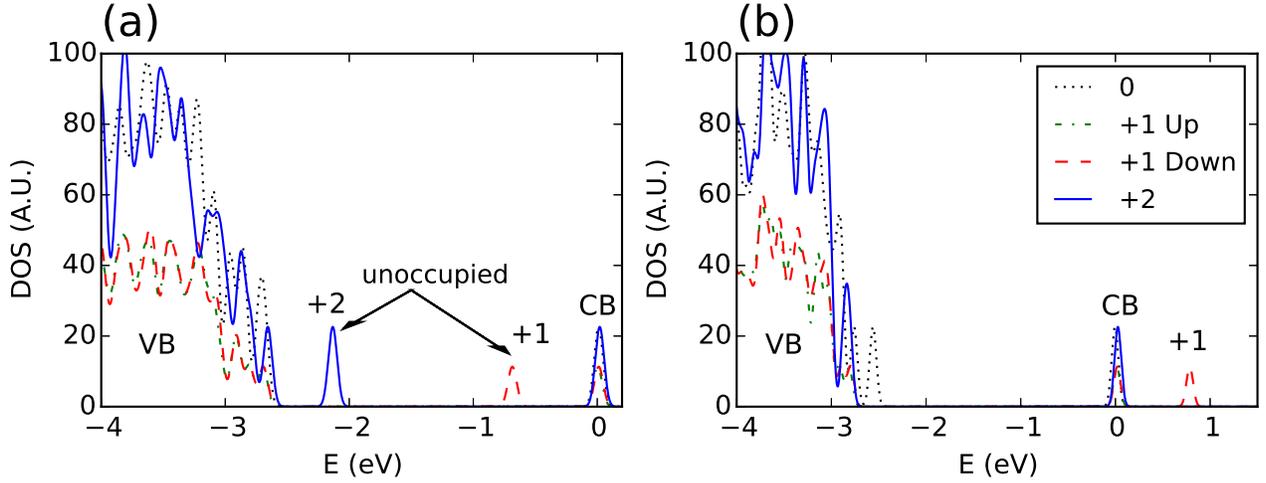}
 \caption{Density of states computed for (a) models number 5 and (b) 8 for the neutral structure (labeled 0), in presence of one (+1) and two positive charges (+2) (holes). The density of states for 1 hole is split in the contributions of the spin up and down. VB and CB label the valence and conduction bands. The two unoccupied states in the band gap in panel (a) are created after the injection of one (+1) or two holes (+2). In panel (b), no localized states are visible in the gap. However, a new localized signature appears above the conduction band whenever a hole is injected.}
 \label{fig:model5DOS}
\end{figure*}

\begin{figure*}[hbtp]
\centering
\includegraphics[scale=1]{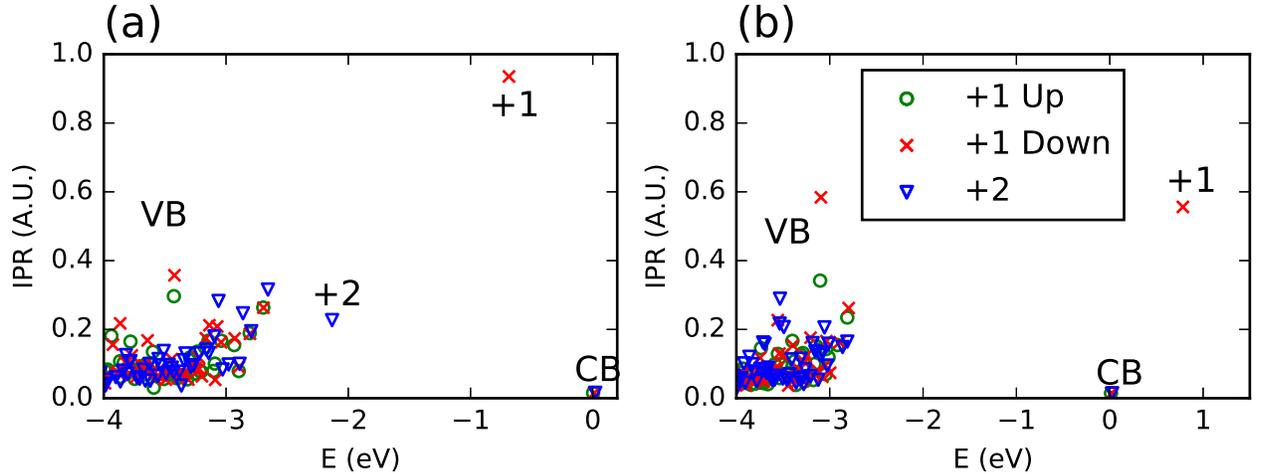}
\label{fig:ipr47}
\caption{IPR plot for models 5 (panel a) and 8 (panel b). The results for one hole (+1 spin up/spin down) is depicted with green dots and red crosses, the results for two holes (+2) are represented with blue triangles. Low IPR values indicate localized states, while a value close to unity hints to the presence of a strongly localized state. In model 8 (panel (b)), no states are formed in the band-gap for two holes due to the creation of a peroxide.}
\end{figure*}

Interestingly, the formation of a polaron destabilizes the tail states in which the hole is injected, moving its signature at higher energies in the gap (figure \ref{fig:model5DOS}). In presence of two holes, if no peroxide is formed, the emptied tail state appears just above the valence band. Conversely, in presence of only one hole, it is generally located close to the conduction band as stressed in figure \ref{fig:model5DOS}(a), or even above it, as in figure \ref{fig:model5DOS}(b). These states are mainly formed from  the signature of anti-bonding oxygen p-orbitals with some minor contributions of the zinc d-orbitals and are therefore similar in terms of composition to those located at the top of the valence band \cite{kamiya2009origins, walsh2009interplay, Nahm2012, Meux2015}. They are nevertheless strongly localized as shown in figure \ref{fig:ipr47}. 

\begin{figure}[hbtp]
\centering
\includegraphics[scale=0.8]{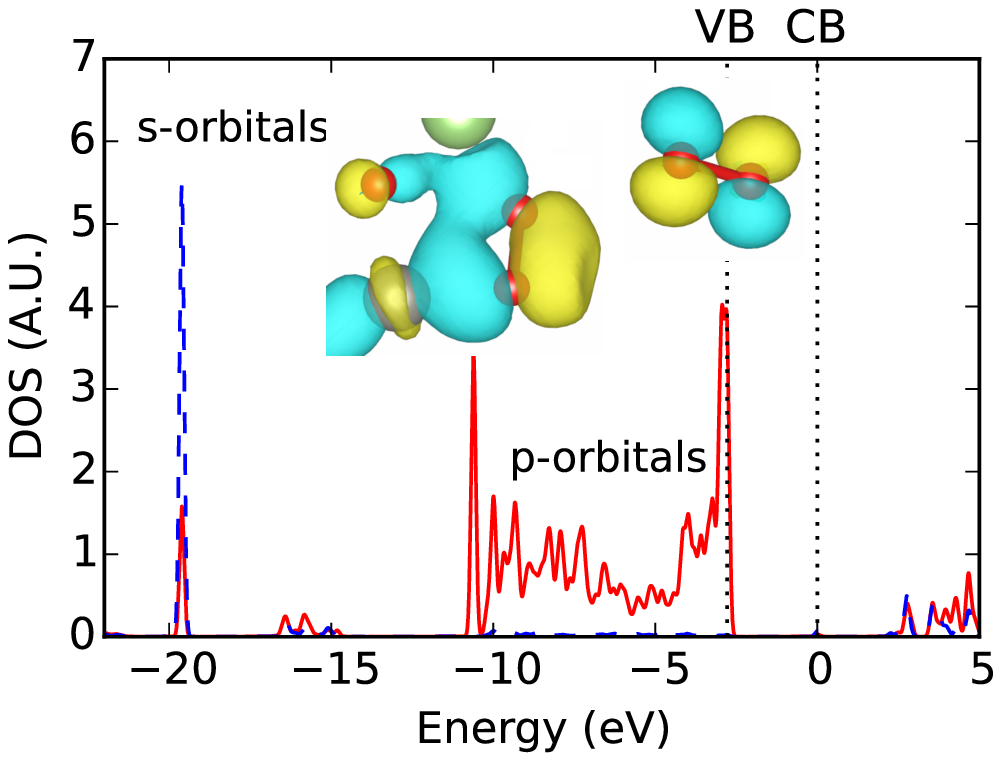}
\caption{Contribution to the density of states of two oxygens forming a peroxide in model 9. The continuous red line provides the contribution of the p-orbitals while the broken blue curve is the one of the s-orbitals. The position of the valence (VB) and of the conduction bands (CB) are shown by two vertical dashed lines. The iso-surface of the wave-function associated to the two major peaks of the p-orbitals DOS, are provided in inset and obtained with Vesta \cite{momma2011vesta}. The left inset shows the $\rm\pi$ bond between the two oxygens of the peroxide while the right one the $\rm\pi^*$ bond.}
\label{fig:contribOO}
\end{figure}

The formation of a peroxide cures these highly localized states at the expense of the formation of a donor. Indeed, whenever a peroxide is formed, no empty state is localized in the band gap. This forces the system to delocalize the remaining electrons required to reach the charge neutrality into the conduction band. The electronic signature of these peroxides is associated with three major electronic contributions: one located deep in the valence band, arises from the s-orbital and has an anti-bonding wave-function between the two oxygen atoms but a bonding one with respect to the neighboring metals. The two other contributions are linked to the p-orbitals through a bonding  $\rm\pi$ state that is located deep in the valence band and an anti-bonding $\rm\pi^*$ one, near the top of the valence band, as illustrated in figure \ref{fig:contribOO}   in agreement with previous results from Nahm et al\cite{Nahm2012}.

\begin{figure}[hbtp]
\centering
\includegraphics[scale=1]{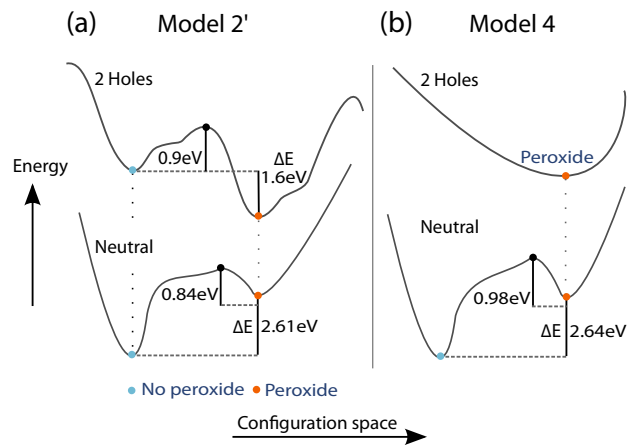}
\caption{ Configuration diagrams for the formation of a peroxide (a) with or (b) without a barrier. The barrier height and $\rm \Delta E$ are provided for the models labeled 2' (panel a) and 4 (panel b). The upper curves show the configuration space in presence of 2 holes while the lower ones are for neutral models.}
\label{fig:path}
\end{figure}

In absence of peroxides, the re-injection of electrons in models that initially own either one or two hole(s), leads to the recovery of the initial atomic configuration. This suggests that holes can easily recombine upon electron injection. In presence of peroxides, however, no recombination is observed. The recovery barriers, provided in table \ref{table:barrier}, range from 0.84 to 1.23eV and are consistent with the experimental value of 1.25eV\citep{HossainChowdhury2013} observed for NBIS. Since original neutral structures (i.e. without peroxide) are more stable than the ones with a peroxide (table \ref{table:Eperox}), these defects are meta-stable. This situation is schematically illustrated in figure \ref{fig:path}(b) for model 4: Whenever 2 holes are injected into the structure, a peroxide is formed spontaneously (as illustrated in the upper curve of figure \ref{fig:path}(b)). However, when electrons are re-injected to neutralize the charge, the peroxide structure remains maintained thanks to its large energy barrier that forbids recovery (see the lower curve of figure \ref{fig:path}(b)). 

\begin{table}
\begin{tabular}{c|c}
Model &  Barrier (eV) \\ 
\hline 
2' & 0.84  \\
4 & 0.98 \\ 
8 & 0.91 \\ 
9 & 1.23 \\ 
Ref \cite{Nahm2012} & 0.9
\end{tabular} 
\caption{Recovery barrier heights for the peroxides in the neutral state.The structure labeled 2' corresponds to the structure where the formation of a peroxide is forced as described in the body of the text.}
\label{table:barrier}
\end{table}

\begin{table}
\begin{tabular}{c|c|c}
Model & $\rm\Delta E (eV)$ & Number of atoms \\ 
\hline 
2' & 2.61 & 105 \\ 
4 & 2.64 & 105 \\ 
8 & 1.60 & 105 \\ 
9 & 2.29 & 105 \\ 
11 & 0.41 & 315 \\ 
Ref \cite{Nahm2012} & 1.25 & 168
\end{tabular} 
\caption{Energy differences between the initial neutral structure and neutral structures with a peroxide. $\rm\Delta E$ is equal to the total energy of the system with the peroxide minus the one without the peroxide. A positive value indicates that the peroxide is meta-stable in the neutral state. The structure labeled 2' corresponds to the structure where the formation of a peroxide is enforced, as described in the body of the text.}
\label{table:Eperox}
\end{table}

The formation of a peroxide upon the addition of two holes was already described by Nahm et al.\cite{Nahm2012} using an atomic model containing 168 atoms. These simulations were performed using a combination of Hubbard corrected DFT and hybrid (HSE) simulations. The peroxide obtained using this approach was also found to be meta-stable with the difference of energy between the system with and without the peroxide of $\rm\Delta E$=1.25eV (table \ref{table:Eperox}) and a barrier of 0.97eV for recovery. Interestingly, these authors also report that a barrier of 0.26eV exists between the original structure in which the holes were injected, and the final one, where the peroxide was formed. This barrier is however not observed for four (out of eleven) of our models which spontaneously formed a peroxide upon charging.

The results of Nahm et al.\cite{Nahm2012} suggest that there should be an energy barrier for the generation of a peroxide in our seven other models. To verify this hypothesis, we enforced the creation of a peroxide in model 1 by moving an oxygen at a distance of ~1.5$\rm \AA$ from another oxygen site. Subsequently, the structure was relaxed in presence of two holes. As expected, the operation results into the creation of a peroxide structure which is more stable than the original one (without the peroxide) by 1.6eV, as schematically represented in figure \ref{fig:path}(a). Nevertheless, in this situation, a large barrier of 0.9eV needs to be overcome for the formation of this peroxide. Upon the re-injection of charges, this new peroxide becomes meta-stable as the original neutral state is 2.61eV more stable, while the resulting barrier for recovery is 0.84eV. Similarly, to the other peroxides (table \ref{table:barrier}), the energetic barrier to be crossed to dissolve the peroxide remains high, indicating that the recovery process is kinetically hindered.”

\begin{figure}[hbtp]
\centering
\includegraphics[scale=1]{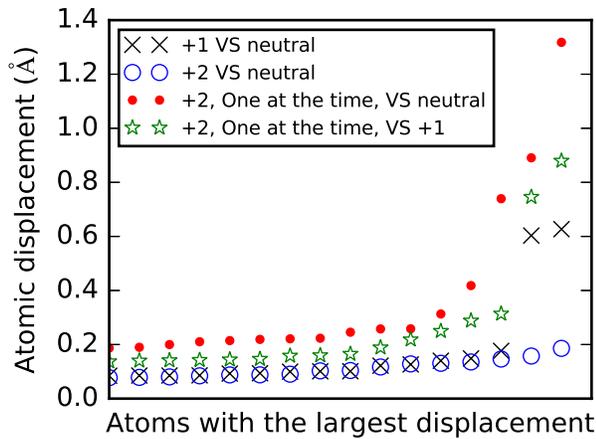}
\caption{Distribution of the atomic displacements due to the injection of one or two positive charges in Model 6. The displacements are sorted in ascending order. A peroxide forms in the model when two positive charges are injected one at the time (red dots and green stars). The largest displacements are obtained for the formation of the peroxide starting from a neutral state (red dots). This displacements is reduced if the formation occurring in the structure already contains one positive charge (green stars).}
\label{fig:displacement}
\end{figure}

The trapping of a single hole is also found to promote the creation of peroxides. Indeed, the results presented are obtained by assuming that a direct injection of one or two holes takes place in the structure. Nevertheless, due to the large structural modifications occurring upon charge injection, one may wonder whether the simultaneous injection of two holes leads to similar results than injecting them one at a time, after atomic relaxation. The results are intuitively expected to be equivalent and the relaxation energy should hence increase with the number of holes injected. This is indeed holding true for most of the models (see table \ref{table:relaxE}). However, model 6 shows a reverse trend: the relaxation energy in presence of two holes (0.26eV) is substantially lower than in presence of one hole (0.9eV). To further investigate this anomaly, we added a second hole to the atomic structure of model number 6 starting from the atomic configuration generated with one hole. The resulting atomic relaxation leads to the formation of a peroxide, which was absent from the initial model relaxed directly in the presence of two holes. As exemplified in figure \ref{fig:displacement}, the atomic displacement required  to create this peroxide from the neutral system is about $\rm 1.3 \AA$. In the present case, the direct injection of two positive charges hardly leads to an atomic displacement of 0.2\AA , explaining the low relaxation energy. However, if the holes are injected one at the time, the peroxide formation occurs in two steps associated with large atomic displacements (up to $\rm \sim 0.9 \AA$). The injection of one hole at a time combined with a subsequent atomic relaxation process therefore favors the formation of peroxides.

We conclude that a hole created in the tail states can either localize into a polaron or mediate the formation of a peroxide. The transport of holes in a-IGZO is therefore particularly difficult, fully corroborating the fact that p-type conductivity could never be observed.  It can however be argued that if holes are generated away from the tail states, i.e. deep in the valence band, they could also be more mobile. At this stage, it is instructive to compare with the behavior reported for other oxides. Interestingly, a pronounced self-trapping process is found to occur in crystalline oxides such as $\rm Ga_2O_3$ and $\rm In_2O_3$\cite{Varley2012} where, by definition, no tail states are present. Additionally, there is no sharp variation between the tail states and the valence band itself, since both share the same composition (i.e. oxygen p-orbitals). It is therefore unlikely that mobile holes get generated at any meaningful energy in a-IGZO, which contributes to explain the low off current reported for a-IGZO transistors. 

\begin{figure}[hbtp]
\centering
\includegraphics[scale=1]{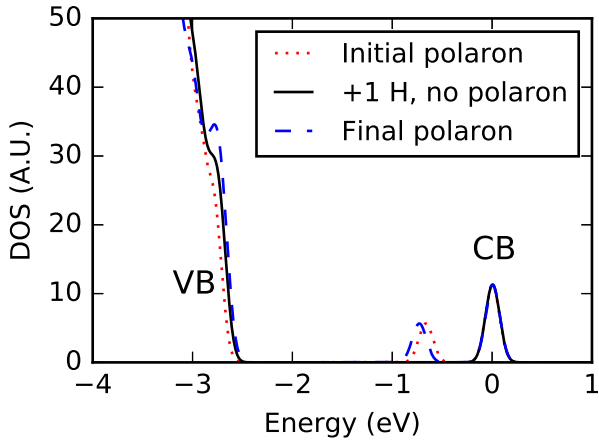}
\caption{Density of states of a model with a polaron before (doted red line) and after (dashed blue line) the addition of an hydrogen. The DOS of the model containing the hydrogen and passivating the polaron is provided by the continuous black line. There is a state in the gap only in presence of a polaron which slightly move in energy after the addition and removal of the hydrogen.}
\label{fig:polaronDos}
\end{figure}

As generated holes are easily trapped by polarons, the hole conduction is limited by their diffusion. It is however challenging to study this diffusion process due to the difficulty to move a polaron in our models. We suspect however that hydrogen could mediate it. It is expected that the presence of an additional hydrogen will ionize (and hence passivate) the polaron by capturing its positive charge. Conceptually, one may therefore consider that the hole moved from the polaron to the hydrogen. If the hydrogen is subsequently removed, the polaron is expected to re-form close to the position of the removed hydrogen (which may differ from the original position). To test this hypothesis, we created 4 models where a hydrogen atom was added close to a polaron. As expected, the additional hydrogen removed the polaron state in all the models without inducing a new localized state in the gap as exemplified in figure \ref{fig:polaronDos}. 

\begin{figure*}[hbtp]
\centering
\includegraphics[scale=0.27]{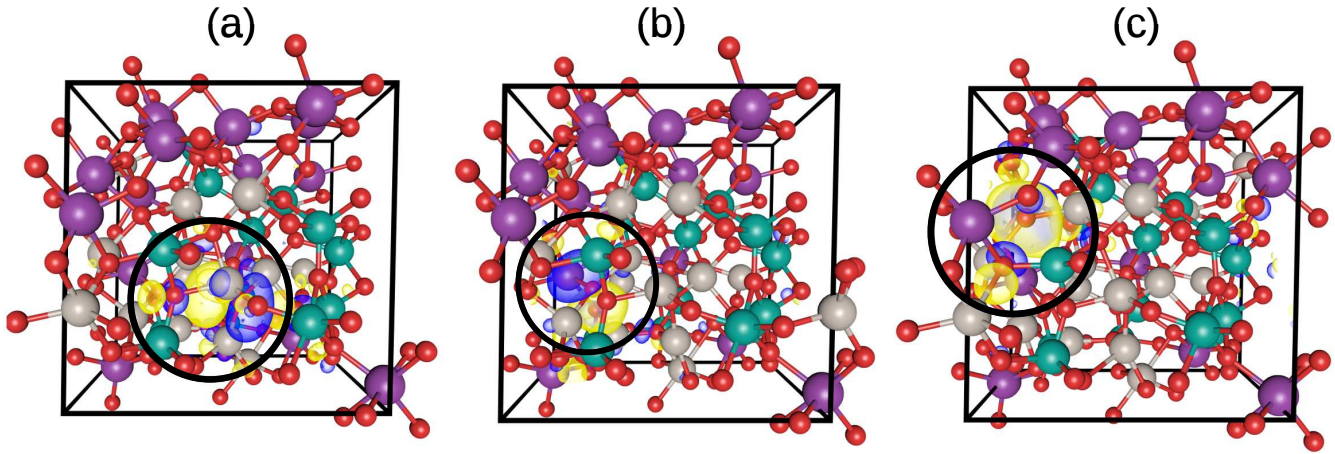}
\caption{Representation of the iso-surface of the wave-function of a polaron state for three different positions of this polaron. The position of the polaron is highlighted by a black circle. The energy path to move the polaron from configuration (a) to (b) or (c) is provided in figure \ref{fig:NEB}. The variations of the position are obtained by adding and removing an hydrogen at different places of the model. The phase of the wave-function is provided by the blue or yellow color of the iso-surface. Oxygen atoms are in red, zinc in grey, Indium in purple and gallium in green. CB and VB indicate the bottom of the conduction band and the top of the valence band.}
\label{fig:polaronpos}
\end{figure*}

\begin{figure}[hbtp]
\centering
\includegraphics[scale=1]{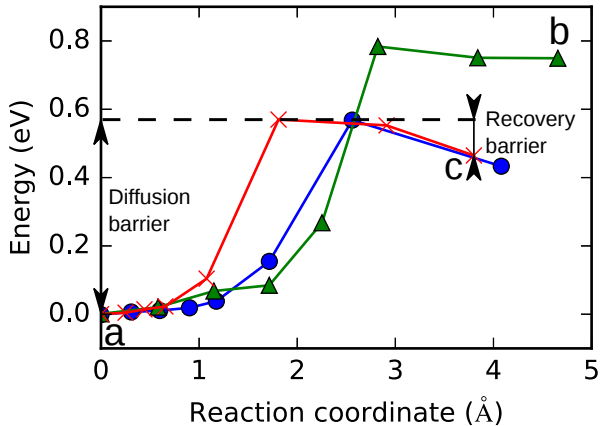}
\caption{Minimum energy path computed for the diffusion of a polaron into three alternative positions. The initial position is labeled (a) and is illustrated in figure \ref{fig:polaronpos}(a). (b) and (c) correspond to two final conformations and are illustrated in figure \ref{fig:polaronpos}(b) and (c). All the final positions lead to a meta-stable state since the process is endoenergetic with respect to the initial configuration.}
\label{fig:NEB}
\end{figure}

Once the hydrogen is removed, the polaron re-forms it-self and in three cases out of four, the position of the polaron changed as localizing itself about 3.2\AA, 4.1\AA or 5.2\AA~ away from its initial position as illustrated in figure \ref{fig:polaronpos}. However, all three polarons now lie in a higher energy configuration, as shown in figure \ref{fig:NEB}. These should hence be considered as being meta-stable transition-states and highlight the difficulty to find another stable position for the polaron in the vicinity of the initial one. From these atomic configurations, we can estimate that the minimum barrier for the diffusion of the polaron is of about 0.6eV, which corresponds to the smallest barrier obtained in figure \ref{fig:NEB}. This barrier is much larger than the thermal energy at room temperature ($\sim$ 0.025eV), which supports a slow diffusion process of these states. Additionally, the activation energy barriers for recovery are small and varying between 0.03 and 0.14eV. They suggest that these polarons can easily back-track to their initial position instead of diffusing further away, supporting again a low diffusion process.

\section{NBIS} 

As described in the introduction, the formation of peroxides was postulated to be at the origin of NBIS in a-IGZO\cite{Nahm2012,Jeong2013b,Robertson2014,Han2015}. Indeed, both the low gate bias and light excitation are expected to induce holes in the tail states, which can translate into the formation of peroxides doping the material. As these are meta-stable whenever the stress is removed, the doping remains resulting in a shift of the transistor curve. 

Nevertheless, NBIS is a slow process \cite{Hung2014,HossainChowdhury2013}. At first sight, this is incompatible with the results obtained. Indeed, the simulations indicate that the formation of a peroxide can be barrier-free, implying that the formation of a peroxide is fast (i.e. in the order of a few atomic vibrations, $\rm \sim 10^{-12}$s) which is contradictory with the pictures of a slow NBIS. Moreover, both light and negative bias should be able to create holes and hence trigger the formation of peroxides. However, in presence of a negative gate potential alone, holes can only be injected from the contacts. Given that holes get trapped in polarons, the hole injection process will be inefficient and likely limited to the regions around the contacts, pinning the Fermi-level. This is expected to seriously restrict the creation of peroxides in the system and therefore the occurrence of NBIS. On the contrary, photons with sufficient energy will create holes in the channel by either a direct excitation of the tail states or of the valence band states. But in absence of a gate bias, the generated electron-hole pairs (excitons) will rapidly recombine, limiting again the creation of peroxides. It results that by combining the two, light can generate excitons in the channel while the bias stress reduces their recombination probability by depleting the semi-conductor from the generated electrons, leaving only the holes which are able to trigger the formation of peroxides. Note that if the Fermi-level is sufficiently close to the valence band, the empty states linked to single holes (found in the upper part of the gap) will lie above the Fermi-level, reducing even further the recombination probability. 

The time dependence of NBIS can be explained by the time required for enough holes to diffuse, or to be generated, into a configuration where they are sufficiently close to each other to trigger the formation of a peroxide. Indeed, although the injection of two holes in a model of 105 atoms represents a very large doping of $\rm \sim 2.6\times10^{21} cm^{-3}$ some models still show barriers for the formation of peroxides. This indicates the existence of a distribution of barriers probably dependent on the quality of the structure. Hence, in high quality a-IGZO, NBIS could then be further slowed-down by the time required to thermally cross these formation barriers.

The role of hydrogen, typically present in high concentration in a-IGZO \cite{Nomura2013} and reported to have a strong negative impact on NBIS \cite{Jeong2013a}, can be explained by its ability to mediate the diffusion of holes as discussed in the results section. Moreover, the presence of a large density of hydrogen must be compensated through the inclusion of additional oxygen atoms\cite{Nomura2013} in order to conserve a carrier density small enough to be able to operate the transistor. As peroxides arise from the interaction between oxygen atoms, the increase of the oxygen concentration in the material will favor their creations, increasing NBIS. 

\section{Conclusion} 

The influence of hole injection on a-IGZO was studied from first-principles calculations. The injection of one hole in the models leads to an increase of the tail states energy and to the localization of the hole in a polaron. Some models show the spontaneous formation of peroxides, which act as electron donors and are meta-stable: they remain in the system even when electrons are re-injected. The localization of holes in polarons contributes to the low off-current and to the difficulty to obtain a p-type conduction in a-IGZO transistors. 

The creation of meta-stable peroxides and the high localization of holes offer a consistent model to explain the origin of NBIS. As holes are trapped into polarons, they cannot be easily moved in the material. This  explains the need of combining the contribution of photons (which generate holes) with a negative gate potential, that stabilizes these holes by depleting the semi-conductor and by pushing the Fermi-level below their respective energy levels. With time, holes diffuse, and lead to the formation of a peroxide that acts as a meta-stable electron donor, shifting the transistor transfer curve.

\section{Supplementary Material} 

See supplementary material for the density of states, structural parameters and formation energies of the different amorphous models presented.

\bibliographystyle{apsrev4-1}
\bibliography{bib}

\end{document}